\documentstyle[12pt]{article}
\input{tcilatex}

\QQQ{Language}{American English}

\begin{document}

\begin{center}
{\bf QUANTUM MECHANICAL PROPAGATORS\medskip\ }

{\bf IN TERMS OF HIDA DISTRIBUTIONS\bigskip\medskip\ \ }

{\bf Angelika Lascheck$^1$}\smallskip\ 

{\bf Peter Leukert$^1$}\smallskip\ 

{\bf Ludwig Streit$^{1,2}$}\smallskip\ 

{\bf Werner Westerkamp$^1$}\medskip\ \ 

{\bf $^1$}BiBoS - Univ. Bielefeld, Bielefeld, Germany\smallskip\ 

{\bf $^2$}Universidade da Madeira, Funchal, Portugal
\bigskip \smallskip\ \ 

Published in Rep. Math. Physics 33 (1993) p221 
\bigskip \smallskip\ \ 

{\bf ABSTRACT}\smallskip\ 
\end{center}

\begin{quotation}
We review some basic notions and results of White Noise Analysis that are
used in the construction of the Feynman integrand as a generalized White
Noise functional. After sketching this construction for a large class of
potentials we show that the resulting Feynman integrals solve the
Schr\"odinger equation .\medskip\bigskip\ \ 
\end{quotation}

{\bf 1. }\underline{{\bf Introduction}}\medskip\ 

The idea of realizing Feynman integrals within the White Noise framework
goes back to \cite{HS}. The ''average over all paths'' is performed with a
Hida distribution as the weight. The existence of such Hida distributions
corresponding to Feynman integrands has been established in \cite{FPS}. In 
\cite{KS} Khandekar and Streit moved beyond the existence theorem by giving
an explicit construction for a large class of potentials including singular
ones. Here we generalize their construction by allowing time-dependent
potentials of noncompact support. Also we prove that the expectation of the
thus constructed Feynman integrand, i.e. the Feynman integral, does indeed
solve the integral equation for quantum mechanical propagators, which
corresponds to the Schr\"odinger equation. Furthermore we elaborate an idea
suggested in \cite{FPS} concerning the physical interpretation of
T-transforms as propagators corresponding to systems with an additional
''source '' term in the potential.\bigskip\ 

\newpage\ {\bf 2. }\underline{{\bf White Noise Analysis}}\medskip\ 

We start from the fundamental Gel'fand triple:\smallskip\ 

$$
S\left( R\right) \subset L^2\left( R\right) \subset S^{\prime }\left(
R\right) , 
$$
\smallskip\ \\where $S^{\prime }\left( R\right) $ denotes the Schwartz
space. Using Minlos' theorem we construct a measure space $\left( S^{\prime
}\left( R\right) ,B,\mu \right) $ called the White Noise space by fixing the
characteristic functional in the following way:\smallskip\ 

$$
C\left( \xi \right) =\int_{S^{\prime }\left( R\right) }\exp i\left\langle
\omega ,\xi \right\rangle \text{ }d\mu \left( \omega \right) =\exp \left(
-\frac 12\left| \xi \right| _0^2\right) \text{ },\text{ }\xi \in S\left(
R\right) ; 
$$
\smallskip\ \\here $\left\langle .,\text{ }.\right\rangle $ denotes the
pairing between $S^{\prime }\left( R\right) $ and $S\left( R\right) $ and $%
\left| .\right| _0$ the norm on $L^2\left( R\right) $.

Within this formalism a version of Wiener's Brownian motion is given by:%
\smallskip\ 

$$
B\left( t\right) :=\left\langle \omega ,1_{\left[ 0,t\right] }\right\rangle
=\int_0^t\omega \left( s\right) \text{ }ds\text{.} 
$$
\smallskip\ 

We now consider the space $\left( L^2\right) $, which is defined to be the
complex Hilbert space $L^2\left( S^{\prime }\left( R\right) ,B,\mu \right) .$
For applications the space $\left( L^2\right) $ is often too small. Here we
enlarge the space $\left( L^2\right) $ by first choosing a special subspace $%
\left( S\right) $ of test functionals. Then the corresponding Gel'fand
triple is:\ 

$$
\left( S\right) \subset \left( L^2\right) \subset \left( S\right) ^{*}. 
$$
\ 

Elements of the space $\left( S\right) ^{*}$ are called Hida distributions
and its explicit construction is given in \cite{HKPS}. Instead of
reproducing this construction here we shall completely characterize Hida
distributions by their $S$- or $T$-transforms $\left( \Phi \in \left(
S\right) ^{*},\text{ }\xi \in S\left( R\right) \right) :$%
\begin{equation}
\label{Tee}T\Phi \left( \xi \right) \equiv \left\langle \left\langle \Phi
,\exp \text{ }i\left\langle .,\xi \right\rangle \text{ }\right\rangle
\right\rangle =\dint_{S^{\prime }\left( R\right) }\exp \left( i\left\langle
\omega ,\xi \right\rangle \right) \Phi \left( \omega \right) d\mu \left(
\omega \right) , 
\end{equation}
$$
S\Phi \left( \xi \right) \equiv \left\langle \left\langle \Phi ,:\exp
\left\langle .,\xi \right\rangle :\right\rangle \right\rangle , 
$$
\ here $\left\langle \left\langle .,.\right\rangle \right\rangle $ denotes
the dual pairing between $\left( S\right) $ and $\left( S\right) ^{*}$ and
we have used the traditional notation:\smallskip\ 

\begin{equation}
\label{Wick}:\exp \left\langle .,\xi \right\rangle :\text{ }\equiv \exp
\left( \left\langle .,\xi \right\rangle -\frac 12\left| \xi \right|
_0^2\right) ,\text{ }\xi \in S\left( R\right) 
\end{equation}
\smallskip\ As $S$- or $T$-transforms of Hida distributions possess analytic
continuations, we have the relation:\smallskip\ 

$$
T\Phi \left( \xi \right) =C\left( \xi \right) \text{ }S\Phi \left( i\xi
\right) . 
$$
\smallskip\ Let us now quote the above mentioned characterization theorem,
which is due to Potthoff and Streit \cite{PS} and has been generalized in
various ways $\left( \text{see eg.\cite{KoS2},\cite{MY},\cite{SW}}\right) .$%
\bigskip\ \ \ 

{\bf Theorem 2.1:} The following statements are equivalent:\ 

\begin{enumerate}
\item  F: $S\left( R\right) \rightarrow C$ such that:\smallskip\ 

(A) ''Ray-Analyticity'': For all $\zeta ,\xi \in S\left( R\right) $ the
mapping $\lambda \mapsto F(\lambda \xi +\zeta ),$ $\lambda \in R,$ extends
to an entire function $z\mapsto f\left( z,\xi ,\zeta \right) ,$ $z$ $\in $ $%
C.$\smallskip\ 

(B) ''Bound'': f is uniformly of order two, i.e. there exist $p\in N_0$ and
constants $P,Q>0$ so that for all $z\in C,$ $\xi \in S\left( R\right) ,$%
\footnote{$\left\{ \left| .\right| p,p\in N\right\} $ is a countable system
of norms topologizing $S\left( R\right) $ . Any of these norms on $S\left(
R\right) $ may be used here.}\ 

$$
\left| f\left( z,\xi ,0\right) \right| \leq P\exp \left( Q\left| z\right|
^2\left| \xi \right| _p^2\right) . 
$$

\item  F is the $S$- transform of a Hida distribution $\Phi \in \left(
S\right) ^{*}.$\ 

\item  F is the $T$- transform of a Hida distribution $\stackrel{\wedge }{%
\Phi }$ $\in \left( S\right) ^{*}.$\ 
\end{enumerate}

\noindent A functional satisfying (A) and (B) is usually called U-functional.%
\smallskip\ 

\noindent Since the space of U-functionals forms an algebra under pointwise
multiplication and $S$- as well as $T$-transform are injective, we can
introduce two algebraic structures in the space $\left( S\right) ^{*}$,
namely the Wick product $\diamond $ and a convolution $*$, which are defined
as follows: Let $\Phi ,\Psi \in \left( S\right) ^{*}$, then\ 

$$
S\left( \Phi \diamond \Psi \right) =S\Phi \cdot S\Psi ,\text{ }\;\,T\left(
\Phi *\Psi \right) =T\Phi \cdot T\Psi . 
$$
\ Now we want to mention two other important consequences of theorem 2.1.
The first one concerns the convergence of sequences of Hida distributions
and can be found in $\cite{HKPS},\cite{PS}.$\bigskip\ \ 

{\bf Theorem 2.2:} Let $\left\{ F_n\right\} _{n\in N}$ denote a sequence of
U-functionals with the following properties:

\begin{enumerate}
\item  For all $\xi \in S\left( R\right) $ , $\left\{ F_n\left( \xi \right)
\right\} _{n\in N}$ is a Cauchy sequence,\smallskip\ 

\item  There exist $P,Q$ $>0$ and $p\in N_0$ , such that the bound\ 
$$
\left| F_n\left( z\xi \right) \right| \leq P\exp \left( Q\left| z\right|
^2\left| \xi \right| _p^2\right) \text{ },\text{ }\xi \in S\left( R\right) , 
\text{ }z\in C 
$$
\smallskip\ holds uniformly in $n\in N$ (here $F_n$ denotes the entire
analytic extension).

Then there is a unique $\Phi \in \left( S\right) ^{*}$ such that $T^{-1}F_n$
converges strongly to $\Phi .$\smallskip\ 
\end{enumerate}

\noindent This theorem is also valid for $S$-transforms. The second
corollary of theorem 2.1 deals with the integration of Hida distributions
which depend on an additional parameter (see$\cite{HKPS},\cite{KS}$).\bigskip%
\ \ 

{\bf Theorem 2.3:} Let $\left( \Omega ,B,m\right) $ denote a measure space
and $\lambda \mapsto \Phi \left( \lambda \right) $ a mapping from $\Omega $
to $\left( S\right) ^{*}$. Let $F\left( \lambda \right) $ denote the $T$%
-transform of $\Phi \left( \lambda \right) $ which satisfies the following
conditions for all $\lambda \in \Omega $:\smallskip\ 

\begin{enumerate}
\item  $\lambda \mapsto F\left( \lambda ,\xi \right) $ is a measurable
function for all $\xi \in S\left( R\right) ,$\smallskip\ 

\item  There exists $p\in N_0$ such that\smallskip%
$$
\ \left| F\text{ }\left( \lambda ,z\xi \right) \right| \leq P\left( \lambda
\right) \exp \left( Q\left( \lambda \right) \left| z\right| ^2\left| \xi
\right| _p^2\right) ,\text{ }\xi \in S\left( R\right) ,\text{ }z\in C 
$$
\smallskip\ with $Q$ $\in $ $L^\infty \left( \Omega ,m\right) $ and $P\in
L^1\left( \Omega ,m\right) .$\smallskip\ 
\end{enumerate}

\noindent Then $\Phi $ is Bochner integrable\footnote{%
with respect to one of the Hilbertian norms topologizing (S)*} and\ 
$$
\int_\Omega \Phi \left( \lambda \right) \text{ }dm\left( \lambda \right) \in
\left( S\right) ^{*} 
$$
\ Let $\varphi \in \left( S\right) $, then\smallskip\ 

$$
\left\langle \left\langle \int_\Omega \Phi \left( \lambda \right) \text{ }%
dm\left( \lambda \right) ,\varphi \right\rangle \right\rangle =\int_\Omega
\left\langle \left\langle \Phi \left( \lambda \right) ,\varphi \right\rangle
\right\rangle \text{ }dm\left( \lambda \right) \text{ .} 
$$
\smallskip\ 

\noindent The last equation allows us to intertwine $T$-transform and
integration\ 
$$
T\left( \int_\Omega \Phi \left( \lambda \right) \text{ }dm\left( \lambda
\right) \right) \left( \xi \right) =\int_\Omega T\left( \Phi \left( \lambda
\right) \right) \left( \xi \right) \,dm\left( \lambda \right) . 
$$
\ 

\noindent Again the same theorem holds for the $S$-transform.\medskip\ \ 

Before we close this section we would like to give two examples of Hida
distributions.\medskip\ 

{\bf Example 2.4:} {\it Donsker}'{\it s $\delta $}-{\it function}\smallskip\ 

Now we study the following informal expression:\smallskip\ 

$$
\Phi =\delta \left( B\left( t\right) -a\right) 
$$

\begin{equation}
\label{Donsker}\Phi =\delta \left( \left\langle .,1_\Delta \right\rangle
-a\right) ,\text{ }a\in R,\text{ }\Delta =\left[ 0,t\right) 
\end{equation}
\smallskip\ The $S$- transform of $\Phi $ is calculated to be \cite{HKPS}$:$%
$$
S\Phi \left( \xi \right) =\frac 1{\sqrt{2\pi t}}\exp \left( -\frac
1{2t}\left( \dint\limits_0^t\xi \left( s\right) \text{ }ds-a\right) ^{{\bf 2}%
}\right) 
$$
\ and theorem 2.1 gives immediately that $\Phi $ is a well defined element
in $\left( S\right) ^{*}.$\medskip\ 

\ {\bf Example 2.5:}\smallskip\ \ 

Let us consider the following informal expression for complex $c\neq \frac
12:$\ 

$$
\exp \left( c\int_a^b\omega ^2\left( s\right) \,ds\right) . 
$$
\ Calculation of its $S$- transform produces a U-functional ''up to an
infinite constant''. So as a renormalization we omit this divergent factor
and get a well defined U- functional:\ 
$$
F\left( \xi \right) =\exp \left( \frac c{1-2c}\int_a^b\xi {}^2\left(
s\right) \,ds\right) 
$$
\ Hence we may define $N\exp \left( c\int_a^b\omega ^2\left( s\right)
\,ds\right) \equiv S^{-1}F$. Then roughly speaking \ 
$$
N\exp \left( c\int_a^b\omega {}^2\left( s\right) \,ds\right) =\frac{\exp
\left( c\int_a^b\omega {}^2\left( s\right) \,ds\right) }{E\left( \exp \left(
c\int_a^b\omega {}^2\left( s\right) \,ds\right) \right) }. 
$$
\ See \cite{HKPS} for all details.\bigskip\smallskip\ \ \ 

{\bf 3. }\underline{{\bf The Feynman integrand as a Hida distribution}}%
\medskip\smallskip\ \ 

We follow Ref.\cite{FPS} and \cite{HS} in viewing the Feynman integral as a
weighted average over Brownian paths. These paths are modeled within the
White Noise framework according to\smallskip\ 
$$
x\left( t\right) \equiv x_0+\sqrt{\frac \hbar m}\int_{t_0}^t\omega \left(
\tau \right) \text{ }d\tau , 
$$
\smallskip\ in the sequel we set $\hbar =m=1.$\smallskip\ 

\noindent \ In Ref.\cite{FPS} the (distribution-valued) weight for the free
motion from $x\left( t_0\right) =x_0$ to\\ $x\left( t\right) =x$ is
constructed from a kinetic energy factor $N\exp \left( \frac
i2\int_{t_0}^t\omega ^2\left( \tau \right) \text{ }d\tau \right) $ and a
Donsker delta function $\delta (x\left( t\right) -x)$. Furthermore a factor $%
\exp \left( \frac 12\text{ }\int_{t_0}^t\omega ^2\left( \tau \right) \text{ }%
d\tau \right) $ is introduced to compensate the Gaussian fall-off of the
White Noise measure in order to mimic Feynman's non-existing ''flat''
measure $D^\infty x.$ Thus in Ref.\cite{FPS} the Feynman integrand for the
free motion reads:\smallskip\ 
$$
J=N\exp \left( \frac{i+1}2\int_{t_0}^t\omega ^2\left( \tau \right) \text{ }%
d\tau \right) \delta \left( x\left( t\right) -x\right) \text{.} 
$$
\smallskip\ However the distribution\smallskip\ 
$$
I_0=N\exp \left( \frac{i+1}2\int_R\omega ^2\left( \tau \right) \text{ }d\tau
\right) 
$$
\smallskip\ has recently been seen to be particulary useful in this context
because of its relation to complex scaling transformation (see Ref.$\cite{S}$%
). It turns out that it is unnecessary to use the time interval $\left[
t_0,t\right] $ in the kinetic energy factor, because the delta function
introduces the interval into the resulting distribution $I_0\delta $. Indeed
it will be shown that $I_0\delta $ produces the correct physical results
(see below especially theorem 3.4). As the choice of $I_0\delta $ rather
than $J$ as a starting point produces only minor modifications in
calculations and formulae, all the pertinent results in Ref.\cite{FPS} and 
\cite{KS} can be established in a completely analogous manner. We give just
a brief account. As in Ref.\cite{FPS} $I_0\delta $ is a Hida distribution,
with $T$- transform given by\ 
$$
TI_0\delta \left( \xi \right) =\frac{\theta \left( t-t_0\right) }{\sqrt{2\pi
i\left| t-t_0\right| }}\exp \left( -\frac i2\left| \xi \right| ^2+\frac
i{2\left| t-t_0\right| }\left( \int_{t_0}^t\xi \left( \tau \right) d\tau
+x-x_0\right) ^2\right) \text{,} 
$$
where we have introduced the Heaviside function to ensure causality.
Furthermore the Feynman integral $E\left( I_0\delta \right) =TI_0\delta
\left( 0\right) $ is indeed the (causal) free particle propagator $\frac{%
\theta \left( t-t_0\right) }{\sqrt{2\pi i\left| t-t_0\right| }}$ $\exp
\left[ \frac i{2\left| t-t_0\right| }\text{ }\left( x-x_0\right) ^2\right] $%
. Not only the expectation but also the $T$- transform has a physical
meaning. By a formal integration by parts\ 
$$
TI_0\delta \left( \xi \right) =E\left( I_0\delta \text{ }e\text{ }%
^{-i\int_{t_0}^tx\left( \tau \right) \stackrel{\cdot }{\xi {}}\left( \tau
\right) \,d\tau }\right) \text{ }e\text{ }^{ix\xi \left( t\right) -ix_0\xi
\left( t_0\right) }\text{ }e\text{ }^{-\frac i2\left| \xi _{\left[
t_0,t\right] }c\right| ^2}. 
$$

\noindent ($\xi _{\left[ t_0,t\right] }c$ denotes the restriction of $\xi $
to the complement of $\left[ t_0,t\right] $). The term $e$ $%
^{-i\int_{t_0}^tx\left( \tau \right) \stackrel{\cdot }{\xi {}}\left( \tau
\right) \,d\tau }$ would thus arise from a time-dependent potential $W\left(
x,t\right) =$ $\stackrel{\cdot }{\xi \text{{}{}}{}}\left( \tau \right) x$.
And indeed it is straightforward to verify that\ 
\begin{equation}
\label{KoDef}TI_0\delta \left( \xi \right) =K_0^{\left( \stackrel{\cdot }{%
\xi }\right) {}}\left( x,t|x_0,t_0\right) \text{ }e\text{ }^{ix\xi \left(
t\right) -ix_0\xi \left( t_0\right) }\text{ }e\text{ }^{-\frac i2\left| \xi
_{\left[ t_0,t\right] }c\right| ^2}, 
\end{equation}
\ where%
$$
K_0^{\left( \stackrel{\cdot }{\xi }\right) }\left( x,t|x_0,t_0\right) =\frac{%
\theta \left( t-t_0\right) }{\sqrt{2\pi i\left| t-t_0\right| }}\;\times 
$$
\begin{equation}
\label{Ko}\exp \left( ix_0\xi \left( t_0\right) -ix\xi \left( t\right)
-\frac i2\left| \xi _{\left[ t_0,t\right] }\right| ^2+\frac i{2\left|
t-t_0\right| }\left( \int_{t_0}^t\xi \left( \tau \right) d\tau +x-x_0\right)
^2\right) 
\end{equation}
\ is the Green's function corresponding to the potential $W,$ i.e. $%
K_0^{\left( \stackrel{\cdot }{\xi }\right) }$ obeys the Schr\"odinger
equation\smallskip\ 
$$
\left( i\partial _t+\frac 12\partial _x^2-\stackrel{\cdot }{\xi {}}\left(
t\right) x\right) K_0^{\left( \stackrel{\cdot }{\xi }\right) }\left(
x,t|x_0,t_0\right) =i\,\delta \left( t-t_0\right) \,\delta \left(
x-x_0\right) . 
$$
\smallskip\ More generally one calculates\ 
\begin{equation}
\label{Tnd}T\left( I_0\stackrel{n+1}{\stackunder{i=1}{\Pi }}\delta \left(
x\left( t_i\right) -x_i\right) \right) \left( \xi \right) =e^{-\frac
i2\left| \xi _{\left[ t_0,t\right] }c\right| ^2}{}e^{\text{ }^{ix\xi \left(
t\right) -ix_0\xi \left( t_0\right) }}\stackrel{n+1}{\stackunder{i=1}{\Pi }}%
K_0^{\left( \stackrel{\cdot }{\xi }\right) }\left(
x_i,t_i|x_{i-1},t_{i-1}\right) . 
\end{equation}
\smallskip
Here $t_0<t_1<...<t_n<t_{n+1}\equiv t$ and $x_{n+1}\equiv x$ .\medskip\ 

In order to pass from the free motion to more general situations, one has to
give a rigorous definition of the heuristic expression\smallskip\ 
$$
I=I_0\delta \exp \left( -i\int_{t_0}^tV\left( x\left( \tau \right) \right) 
\text{ }d\tau \right) . 
$$
\smallskip\ 

In Ref.\cite{KS} Khandekar and Streit accomplished this by pertubative
methods in case $V$ is a finite signed Borel measure with compact support.
Here we give a brief summary of the construction taking into account the
afore-mentioned modification. Also we generalize the construction by
allowing time-dependent potentials and a Gaussian fall-off instead of a
bounded support. Let $\Delta \equiv \left[ {\sf T_0},{\sf T}\right] \supset
[t_0,t]$ and let $v$ be a finite signed Borel measure on $R\times \Delta $ .
Let $v_x$ denote the marginal measure\ 
$$
v_x\left( A\subset B\left( R\right) \right) \equiv v\left( A\times \Delta
\right) 
$$
similary\ 
$$
v_t\left( B\subset B\left( \Delta \right) \right) \equiv v\left( R\times
B\right) . 
$$
\ 

We assume that $v_x$ and $v_t$ satisfy:

i ) $\exists \,R>0$ $\forall \,r>R:\left| v_x\right| \left( \left\{ x:\text{ 
}\left| x\right| >r\right\} \right) <e^{-\beta r^2}$ for some $\beta >0$ ,

ii ) $\left| v_t\right| $ has a $L^\infty $density.\smallskip\ 

\noindent Let us first describe heuristically the construction by treating $%
v $ as an ordinary function $V$ before stating the rigorous result 3.1. The
starting point is a power series expansion of $\exp \left(
-i\int_{t_0}^tV\left( x\left( \tau \right) ,\tau \right) d\tau \right) $
using $V\left( x\left( \tau \right) ,\tau \right) =\int dx\,V\left( x,\tau
\right) \,\delta \left( x\left( \tau \right) -x\right) :$

$$
\exp \left( -i\int_{t_0}^tV\left( x\left( \tau \right) ,\tau \right) d\tau
\right) =\dsum\limits_{n=0}^\infty \left( -i\right) ^n\int_{\Delta _n}d^nt%
\stackrel{n}{\stackunder{i=1}{\Pi }}\int dx_i\,V\left( x_i,t_i\right) \delta
\left( x\left( t_i\right) -x_i\right) 
$$
\smallskip\ where $\Delta _n=\left\{ \left( t_1,...,t_n\right)
|\,t_0<t_1<...<t_n<t\right\} $.\bigskip\ \ 

{\bf Theorem 3.1}:%
$$
I=I_0\delta \left( x\left( t\right) -x\right) +\dsum\limits_{n=1}^\infty
\left( -i\right) ^n\int_{R^n}\int_{\Delta _n}\stackrel{n}{\stackunder{i=1}{%
\Pi }}v\left( dx_i,dt_i\right) \,I_0\delta \left( x\left( t\right) -x\right) 
\stackrel{n}{\stackunder{j=1}{\Pi }}\delta \left( x\left( t_j\right)
-x_j\right) 
$$
\smallskip exists as a Hida distribution in case $V$ obeys i ) and ii ). 
\medskip\ 

{\bf Sketch of Proof:}\medskip\ \ 

{\bf 1)} $I_n=\int_{R^n}\int_{\Delta _n}\stackrel{n}{\stackunder{i=1}{\Pi }}%
v\left( dx_i,dt_i\right) \,I_0\delta \left( x\left( t\right) -x\right) 
\stackrel{n}{\stackunder{j=1}{\Pi }}\delta \left( x\left( t_j\right)
-x_j\right) $ is a Hida distribution for $n\geq 1$. This is shown by
applying theorem 2.3. Choose $q$ $>2$ and 0$<\gamma <\beta /q$ and $p$ such
that $\frac 1p+\frac 1q=1.$ Formulae (\ref{Ko}) (\ref{Tnd}) yield the
estimate%
$$
\left| T\left( I_0\stackrel{n+1}{\stackunder{j=1}{\Pi }}\delta \left(
x\left( t_j\right) -x_j\right) \right) \left( z\xi \right) \right| 
$$
$$
\leq \left( \stackrel{n+1}{\stackunder{i=1}{\Pi }}\frac{\Theta (t_i-t_{i-1}) 
}{\sqrt{2\pi \left| t_i-t_{i-1}\right| }}\right) \exp \left( \gamma \left( 
\stackunder{0\leq j\leq n+1}{\sup }\left| x_i\right| \right) ^2\right) \exp
\left( \frac 12(1+\frac 2\gamma +|\Delta |{\sf )|}z{\sf |^2}\left| \xi
\right| _s^2\right) \text{,} 
$$
where $s$ is such that $\stackunder{t\in \Delta }{\sup }\left| \xi \left(
t\right) \right| <\left| \xi \right| _s$. The property i) of $v$ yields that%
\\ $e^{\gamma x^2}\in L^q(R\times \Delta ,\left| v\right| )$. Let $Q\equiv
\left( \int_R\left| v_x\right| (dx)e^{\gamma qx^2}\right) ^{\frac 1q}$, then%
$$
\left( \int_{R^n}\int_{\Delta ^n}\stackrel{n}{\stackunder{i=1}{\Pi }}\left|
v\right| (dx_i,dt_i)e^{\gamma q\left( \stackunder{0\leq i\leq n+1}{\sup }%
\left| x_i\right| \right) ^2}\right) ^{\frac 1q}\leq e^{\gamma \left|
x_0\right| ^2}e^{\gamma \left| x\right| ^2}Q^n. 
$$
Using the property ii) of $v$ and the formula%
$$
\int_{\Delta _n}d^nt\stackrel{n+1}{\stackunder{j=1}{\Pi }}\frac 1{(2\pi
\left| t_j-t_{j-1}\right| )^\alpha }=\left( \frac{\Gamma (1-\alpha )}{(2\pi
)^\alpha }\right) ^{n+1}\frac{\left| t-t_0\right| ^{n(1-\alpha )-\alpha }}{%
\Gamma \left( (n+1)(1-\alpha )\right) },\text{ }\alpha <1 
$$
we obtain the following estimate:%
$$
\left( \int_{R^n}\int_{\Delta _n}\stackrel{n}{\stackunder{i=1}{\Pi }}\left|
v\right| \left( dx_i,dt_i\right) \stackrel{n+1}{\stackunder{j=1}{\Pi }}%
\left( \frac 1{\sqrt{2\pi \left| t_j-t_{j-1}\right| }}\right) ^p\right)
^{\frac 1p}\leq \,\left| v_t\right| _\infty ^{\frac np}\frac{\Gamma (\frac{%
2-p}2)^{\frac{n+1}p}}{(2\pi )^{\frac{n+1}2}}\frac{|\Delta |^{\frac np-\frac
12(n+1)}}{\Gamma \left( (n+1)(\frac{2-p}2)\right) ^{\frac 1p}} 
$$
Let%
$$
C_n(x,|\Delta |)\equiv e^{\gamma \left| x_0\right| ^2}e^{\gamma \left|
x\right| ^2}Q^n\left| v_t\right| _\infty ^{\frac np}\frac{\Gamma (\frac{2-p}%
2)^{\frac{n+1}p}}{(2\pi )^{\frac{n+1}2}}\frac{|\Delta |^{\frac np-\frac
12(n+1)}}{\Gamma \left( (n+1)(\frac{2-p}2)\right) ^{\frac 1p}}\text{ .} 
$$
H\"older's inequality yields the following estimate:%
$$
\left| \int_{R^n}\int_{\Delta _n}\stackrel{n}{\stackunder{i=1}{\Pi }}v\left(
dx_i,dt_i\right) T\left( I_0\stackrel{n+1}{\stackunder{j=1}{\Pi }}\delta
\left( x\left( t_j\right) -x_j\right) \right) \left( z\xi \right) \right| 
$$
\begin{equation}
\label{Bound1}\leq C_n\exp \left( \frac 12(1+\frac 2\gamma +|\Delta |{\sf )|}
\text{z}{\sf |^2}\left| \xi \right| _s^2\right) 
\end{equation}
\ This establishes the bound required for the application of theorem 2.3 and
hence $I_n$ exists as a Bochner integral in $\left( S\right) ^{*}$.\smallskip%
\ 

{\bf 2)} $I=\dsum\limits_{n=0}^\infty I_n$ exists in $\left( S\right) ^{*}.$%
\smallskip\ 

As the $C_n$ are rapidly decreasing in $n$ the hypotheses of theorem 2.2 are
fulfilled and hence the convergence in $\left( S\right) ^{*}$ is established.%
\marginpar{$\Box $}\medskip\\As an example of the class of admissible
potentials take any finite signed Borel measure $v$ satisfying i) on $R$.
This can be as singular as desired, e.g. a sum of Delta's such as $%
\dsum\limits_{n\in N}e^{-n^2}\delta _n$ or a devil' s staircase. Now take
two bounded measurable functions f and g on $\Delta $. Use one to move the
potential around and the other one to vary its strength: $%
v(x,t)=f(t)\,v(x-g(t))$.\smallskip\ \medskip\ \ \ 

As in the case of the free motion we expect\ 
\begin{equation}
\label{KDef}K^{\left( \stackrel{\cdot }{\xi }\right) }\left(
x,t|x_0,t_0\right) \equiv \text{ }e\text{ }^{+\frac i2\left| \xi _{\left[
t_0,t\right] }c\right| ^2}\text{ }e\text{ }^{-ix\xi \left( t\right) +ix_0\xi
\left( t_0\right) }\,TI\left( x,t|x_0,t_0\right) \left( \xi \right) 
\end{equation}
\ to be the propagator corresponding to the potential $W\left( x,t\right)
=V\left( x,t\right) +\stackrel{\cdot }{\xi }\left( t\right) x$. More
precisely we have to use the measure $\omega \left( dx,dt\right) =v\left(
dx,dt\right) +\stackrel{\cdot }{\xi }\left( t\right) x$ $dx\ dt$.\smallskip\ 

We now proceed to show some properties of $K^{\left( \stackrel{\cdot }{\xi }%
\right) }$. As the propagators $K_0^{\left( \stackrel{\cdot }{\xi }\right) }$
are continuous on R$^2\times \Delta _2$ (see (\ref{Ko})), the product $%
\stackrel{n+1}{\stackunder{j=1}{\Pi }}K_0^{\left( \stackrel{\cdot }{\xi }%
\right) }\left( x_j,t_j|x_{j-1},t_{j-1}\right) $ is continuous on $%
R^{n+1}\times \Delta _{n+1}$.\ Set 
\begin{equation}
\label{Series}K^{\left( \stackrel{\cdot }{\xi }\right) }\left(
x,t|x_0,t_0\right) =\dsum\limits_{n=0}^\infty K_n^{\left( \stackrel{\cdot }{%
\xi }\right) }\left( x,t|x_0,t_0\right) 
\end{equation}
where\ 
$$
K_n^{\left( \stackrel{\cdot }{\xi }\right) }\left( x,t\mid x_0,t_0\right)
=\left( -i\right) ^n\int_{R^n}\int_{\Delta _n}\stackrel{n}{\stackunder{i=1}{%
\Pi }}v\left( dx_i,dt_i\right) \stackrel{n+1}{\stackunder{j=1}{\Pi }}%
K_0^{\left( \stackrel{\cdot }{\xi }\right) }\left(
x_j,t_j|x_{j-1},t_{j-1}\right) \text{.} 
$$
As the test functions $\xi $ are real the explicit formula (\ref{Ko}) yields 
\begin{equation}
\label{bound2}|K_0^{\left( \stackrel{\cdot }{\xi }\right) }\left(
x,t|x_0,t_0\right) |=\frac{\Theta (t-t_0)}{\sqrt{2\pi |t-t_0|}}\equiv M_0 
\end{equation}
and for n$\geq 1$ the bounds\ 
\begin{equation}
\label{bound3}\left| K_n^{\left( \stackrel{\cdot }{\xi }\right) }\left(
x,t|x_0,t_0\right) \right| \leq \int_{R^n}\int_{\Delta _n}\stackrel{n}{%
\stackunder{i=1}{\Pi }}|v|\left( dx_i,dt_i\right) \stackrel{n+1}{\stackunder{%
j=1}{\Pi }}\frac 1{\sqrt{2\pi \left| t_j-t_{j-1}\right| }} 
\end{equation}
$$
\leq |v_t|_\infty ^n\frac{|t-t_0|^{\frac{n-1}2}}{2^{\frac{n+1}2}\Gamma ( 
\frac{n+1}2)}\leq |v_t|_\infty ^n\frac{|\Delta |^{\frac{n-1}2}}{2^{\frac{n+1}%
2}\Gamma (\frac{n+1}2)}\equiv M_n\text{.} 
$$
\ Hence $\stackrel{n+1}{\stackunder{j=1}{\Pi }}K_0^{\left( \stackrel{\cdot }{%
\xi }\right) }\left( x_j,t_j|x_{j-1},t_{j-1}\right) $ is integrable on $%
R^n\times \Delta _n$ with respect to $v^n$. (This is also established in the
course of a detailed proof of theorem 3.1 and we have reproduced the
argument here for the convenience of the reader.) Thus we can apply Fubini's
theorem to change the order of integration in $K_n^{\left( \stackrel{\cdot }{%
\xi }\right) }$\ to obtain%
$$
K_n^{\left( \stackrel{\cdot }{\xi }\right) }\left( x,t|x_0,t_0\right)
=-i\iint v\left( dx_n,dt_n\right) K_0^{\left( \stackrel{\cdot }{\xi }\right)
}\left( x,t|x_n,t_n\right) \text{ }\times 
$$
$$
\left( -i\right) ^{n-1}\int_{R^{n-1}}\int_{\Delta _{n-1}^{^{\prime }}}%
\stackrel{n-1}{\stackunder{i=1}{\Pi }}v\left( dx_i,dt_i\right) \stackrel{n}{%
\stackunder{j=1}{\Pi }}K_0^{\left( \stackrel{\cdot }{\xi }\right) }\left(
x_j,t_j|x_{j-1},t_{j-1}\right) 
$$
$(\Delta _{n-1}^{^{\prime }}=\left\{ (t_1,...,t_{n-1})\mid
\,t_0<t_1<...<t_{n-1}<t_n\right\} )$. This establishes the following
recursion relation for $K_n^{\left( \stackrel{\cdot }{\xi }\right) }$\ 
\begin{equation}
\label{Rec}K_n^{\left( \stackrel{\cdot }{\xi }\right) }\left(
x,t|x_0,t_0\right) =\left( -i\right) \iint v\left( dy,d\tau \right)
K_0^{\left( \stackrel{\cdot }{\xi }\right) }\left( x,t|y,\tau \right)
K_{n-1}^{\left( \stackrel{\cdot }{\xi }\right) }\left( y,\tau
|x_0,t_0\right) \text{.} 
\end{equation}
We now claim that the series $K^{\left( \stackrel{\cdot }{\xi }\right)
}\left( y,\tau |x_0,t_0\right) =\dsum\limits_nK_n^{\left( \stackrel{\cdot }{%
\xi }\right) }\left( y,\tau |x_0,t_0\right) $ converges uniformly in $y,\tau 
$ on $R\times \left( t_0,{\sf T}\right) $. To see this recall the above
estimate (\ref{bound3}) which is uniform in $y,\tau .$ Because the $M_n$ are
rapidly decreasing it follows that\ 
$$
\stackrel{\infty }{\dsum\limits_{n=1}}\sup \left\{ \left| K_n^{\left( 
\stackrel{\cdot }{\xi }\right) }\left( y,\tau |x_0,t_0\right) \right| ;\text{%
(}y,\text{ }\tau )\in R\times \left( t_0,{\sf T}\right) \text{ }\right\}
\leq \stackrel{\infty }{\dsum\limits_{n=1}}\text{ }M_n<\infty \text{ .} 
$$
Due to the uniform convergence we may interchange summation and integration
in the following expression\ 
$$
-i\iint v\left( dy,d\tau \right) K_0^{\left( \stackrel{\cdot }{\xi }\right)
}\left( x,t|y,\tau \right) K^{\left( \stackrel{\cdot }{\xi }\right) }\left(
y,\tau |x_0,t_0\right) 
$$
$$
=-i\iint v\left( dy,d\tau \right) K_0^{\left( \stackrel{\cdot }{\xi }\right)
}\left( x,t|y,\tau \right) \dsum\limits_nK_n^{\left( \stackrel{\cdot }{\xi }%
\right) }\left( y,\tau |x_0,t_0\right) 
$$
$$
=\dsum\limits_n-i\iint v\left( dy,d\tau \right) K_0^{\left( \stackrel{\cdot 
}{\xi }\right) }\left( x,t|y,\tau \right) K_n^{\left( \stackrel{\cdot }{\xi }%
\right) }\left( y,\tau |x_0,t_0\right) . 
$$
By the above recursion relation (\ref{Rec}) for $K_n^{\left( \stackrel{\cdot 
}{\xi }\right) }$ this equals\ 
$$
\dsum\limits_nK_{n+1}^{\left( \stackrel{\cdot }{\xi }\right) }\left(
x,t|x_0,t_0\right) =K^{\left( \stackrel{\cdot }{\xi }\right) }\left(
x,t|x_0,t_0\right) -K_0^{\left( \stackrel{\cdot }{\xi }\right) }\left(
x,t|x_0,t_0\right) \text{.} 
$$
\smallskip\ Hence we obtain the following\bigskip\ \ 

{\bf Theorem 3.2:}\medskip\ \ \ 

$K^{\left( \stackrel{\cdot }{\xi }\right) }$ as defined in (\ref{KDef})
obeys the following integral equation:\ 
$$
K^{\left( \stackrel{\cdot }{\xi }\right) }\left( x,t|x_0,t_0\right)
=K_0^{\left( \stackrel{\cdot }{\xi }\right) }\left( x,t|x_0,t_0\right)
-i\iint v\left( dy,d\tau \right) K_0^{\left( \stackrel{\cdot }{\xi }\right)
}\left( x,t|y,\tau \right) K^{\left( \stackrel{\cdot }{\xi }\right) }\left(
y,\tau |x_0,t_0\right) \text{.} 
$$
\ In particular the Feynman integral $E\left( I\right) \equiv K$ obeys the
well-known propagator equation:\ 
$$
K\left( x,t|x_0,t_0\right) =K_0\left( x,t|x_0,t_0\right) -i\iint v\left(
dy,d\tau \right) K_0\left( x,t|y,\tau \right) K\left( y,\tau |x_0,t_0\right) 
\text{.} 
$$
\smallskip\ We now proceed to show that this corresponds to the
Schr\"odinger equation. To prove this we first prepare the following\bigskip%
\ \ 

{\bf Lemma 3.3:}\medskip\ 

The mapping $\left( x,t\right) \mapsto K^{\left( \stackrel{\cdot }{\xi }%
\right) }\left( x,t|x_0,t_0\right) $ is continuous on $R\times \left( t_0,%
{\sf T}\right) $.\medskip

\noindent {\bf Proof}:

Because the series (\ref{Series}) converges uniformly it is sufficient to
show the continuity of $K_n^{\left( \stackrel{\cdot }{\xi }\right) }$. For $%
n=0,1$ this is straightforward from the explicit formula (\ref{Ko}). For $%
n>1 $ we use (\ref{Rec}) and the estimate (\ref{bound3}) to obtain\ 
$$
\left| K_n^{\left( \stackrel{\cdot }{\xi }\right) }\left( x^{\prime
},t^{\prime }|x_0,t_0\right) -K_n^{\left( \stackrel{\cdot }{\xi }\right)
}\left( x,t|x_0,t_0\right) \right| 
$$
$$
\leq M_{n-1}\int_R\int_\Delta \left| v\right| \left( dx_n,dt_n\right) \left|
K_0^{\left( \stackrel{\cdot }{\xi }\right) }\left( x^{\prime },t^{\prime
}|x_n,t_n\right) -K_0^{\left( \stackrel{\cdot }{\xi }\right) }\left(
x,t|x_n,t_n\right) \right| 
$$
\smallskip. Using the explicit form (\ref{Ko}) of $K_0^{\left( \stackrel{%
\cdot }{\xi }\right) }$it is now straightforward to check that%
$$
\int_R\int_\Delta \left| v\right| \left( dx_n,dt_n\right) \left| K_0^{\left( 
\stackrel{\cdot }{\xi }\right) }\left( x^{\prime },t^{\prime
}|x_n,t_n\right) -K_0^{\left( \stackrel{\cdot }{\xi }\right) }\left(
x,t|x_n,t_n\right) \right| 
$$
$$
\leq \text{ }\mid x-x^{\prime }{}\mid C(x,t)\text{ }+\text{ }\mid
t-t^{\prime }\mid ^\alpha C_\alpha (x,t) 
$$
where $0<\alpha <\frac 12$ and $x>x^{\prime }$, $t>t^{\prime }$.%
\marginpar{$\Box $}\bigskip\ \medskip\ 

An application of lemma 3.3 combined with the estimate (\ref{bound2}) shows
that \\$K^{\left( \stackrel{\cdot }{\xi }\right) }\left( .,.|x_0,t_0\right) $
is locally integrable on $R\times \left( {\sf T}_0,{\sf T}\right) $ with
respect to both $v$ and Lebesgues measure. We can thus regard $K^{\left( 
\stackrel{\cdot }{\xi }\right) }$ as a distribution on $D\left( \Omega
\right) \equiv D\left( R\times \left( {\sf T_0},{\sf T}\right) \right) $:

$$
\left\langle K^{\left( \stackrel{\cdot }{\xi }\right) },\varphi
\right\rangle =\iint dx\,dt\,K^{\left( \stackrel{\cdot }{\xi }\right)
}\left( x,t|x_0,t_0\right) \varphi \left( x,t\right) ,\text{ }\varphi \in
D\left( \Omega \right) . 
$$
And we can also define a distribution $vK^{\left( \stackrel{\cdot }{\xi }%
\right) }$ by setting\ 

$$
\left\langle vK^{\left( \stackrel{\cdot }{\xi }\right) },\varphi
\right\rangle =\iint v(dx,\,dt)\,K^{\left( \stackrel{\cdot }{\xi }\right)
}\left( x,t|x_0,t_0\right) \varphi \left( x,t\right) ,\text{ }\varphi \in
D\left( \Omega \right) . 
$$
\ ($K^{\left( \stackrel{\cdot }{\xi }\right) }$ is locally integrable with
respect to $v$, $\varphi $ is bounded with compact support and $v$ is
finite, hence $\varphi K^{\left( \stackrel{\cdot }{\xi }\right) }$ is
integrable with respect to $v)$.

We now proceed to show that $K^{\left( \stackrel{\cdot }{\xi }\right) }$
solves the Schr\"odinger equation as a distribution. To abbreviate we set $%
\hat L=\left( i\partial _t+\frac 12\partial _x^2-\stackrel{\cdot }{\xi }%
\left( t\right) x\right) $ and let $\hat L^{*}$ denote its adjoint. Let $%
\varphi \in D\left( \Omega \right) $. By theorem 3.2 we have\ 

$$
\left\langle \hat LK^{\left( \stackrel{\cdot }{\xi }\right) },\varphi
\right\rangle =\left\langle K_0^{\left( \stackrel{\cdot }{\xi }\right)
}\left( x,t|x_0,t_0\right) -i\iint v\left( dy,d\tau \right) K_0^{\left( 
\stackrel{\cdot }{\xi }\right) }\left( x,t|y,\tau \right) K^{\left( 
\stackrel{\cdot }{\xi }\right) }\left( y,\tau |x_0,t_0\right) ,\hat
L^{*}\varphi \right\rangle . 
$$
\ By Fubini's theorem this equals\ 
$$
\left\langle K_0^{\left( \stackrel{\cdot }{\xi }\right) },\hat L^{*}\varphi
\right\rangle -i\iint v\left( dy,d\tau \right) \left[ \iint
dx\,dt\,K_0^{\left( \stackrel{\cdot }{\xi }\right) }\left( x,t|y,\tau
\right) \hat L^{*}\varphi \left( x,t\right) \right] K^{\left( \stackrel{%
\cdot }{\xi }\right) }\left( y,\tau |x_0,t_0\right) . 
$$
\ 

As $K_0^{\left( \stackrel{\cdot }{\xi }\right) }$ is a Green's function of $%
\hat L$ we obtain\ 
$$
i\varphi \left( x_0,t_0\right) +\iint v\left( dy,d\tau \right) \varphi
\left( y,\tau \right) K^{\left( \stackrel{\cdot }{\xi }\right) }\left(
y,\tau |x_0,t_0\right) =\left\langle i\delta _{x_0}\delta _{t_0},\varphi
\right\rangle +\left\langle vK^{\left( \stackrel{\cdot }{\xi }\right)
},\varphi \right\rangle . 
$$
Hence we have the following\bigskip\ \ 

{\bf Theorem 3.4:}\medskip\ 

$K^{\left( \stackrel{\cdot }{\xi }\right) }$is a Green's function for the
full Schr\"odinger equation, i.e.\ 
$$
\left( i\,\partial _t+\frac 12\partial _x^2-\stackrel{\cdot }{\xi }\left(
t\right) x-v\right) K^{\left( \stackrel{\cdot }{\xi }\right) }\left(
x,t|x_0,t_0\right) =i\,\delta _{x_0}\,\delta _{t_0}. 
$$
In particular the Feynman integral $E\left( I\right) =K$ solves the
Schr\"odinger equation\ 
$$
i\,\partial _t\text{ }K\left( x,t|x_0,t_0\right) =\left( -\frac 12\partial
_x^2+v\right) K\left( x,t|x_0,t_0\right) \text{, for }t>t_0. 
$$
Hence the construction proposed by Khandekar and Streit yields a
(mathematically) rigorously defined Feynman integrand whose expectation is
the correct quantum mechanical propagator.

\bigskip\ 

{\bf Acknowledgements}\medskip\ 

We would like to thank A. Boukricha, M. de Faria and Y. Kondratiev for
various helpful discussions. We are indebted to the University of Madeira
for the hospitality and support extended to us. This work was made possible
by financial support from STRIDE.\bigskip\

\end{document}